\documentstyle[12pt,psfig]{article}
\begin{document}
\date{}                  
\def\br{{\bf r}}
\def\DF{\Delta F}
\def\bk{{\bf k}}
\def\bR{{\bf R}}
\def\<{\langle}
\def\>{\rangle}

\title{Long Range Moves for High Density Polymer Simulations}
\author{J.M.Deutsch\\
University of California, Santa Cruz, U.S.A.}
\maketitle
\abstract{Monte Carlo simulations of proteins are hindered by
the system's high density which often makes local moves ineffective.
Here we devise and test a set of long range moves that work well
even when all sites of a lattice simulation are filled. We demonstrate
that for a 27-mer cube, the ground state of random heteropolymers
can quickly be reached. We discuss results for 48-mer systems
where the ground state is known exactly. For ten sequences that
were examined, the introduction of long range moves speeds up
the search for the ground state by about one order of magnitude.
The method is compared to a fast folding
chain growth algorithm that had previously been used with
much success. The  new algorithm here appears to be more
efficient. The point is illustrated by the folding of an
80-mer four-helix bundle considered previously.}

\section{Introduction}

In recent years attempts to determine the structure of a protein from
its sequence has attracted a great deal of attention. Given a
polypeptide chain, and the correct intra-atomic forces, how can one find
the thermodynamically stable state of the molecule?  It was recognized
early on\cite{levinthal} that this might be a difficult problem owing to
the exponentially large volume of phase space that, apparently, the
molecule typically needs to explore. Further work has born out the
conclusion that the time required to find a solution is very long
indeed. From the point of view of a physicists, one calls the dynamics
of a protein ``glassy"\cite{bryngelson}. From the vantage point of  a
computer scientist, some protein folding models have been shown\cite{unger}, 
and
many are believed  to be "NP complete", which means that even optimal
algorithms folding proteins are expected to depend exponentially on the
chain length.

Even for highly simplified lattice models of proteins, it appears
difficult to fold even quite short chains; however for a particular choice
of interactions, the so-called HP model\cite{lau}, an efficient folding
algorithm exists\cite{yue} that can fold chains as long as 80 monomers.
This method is quite specific to this model, and the commonly used  but more
general Monte Carlo  method of local chain movement based on the Metropolis 
algorithm\cite{metropolis,verdier,hilhorst}, 
has not been able to fold even 48-mers.
A recent paper on the design of the HP model\cite{whoops} illustrates 
this point. A group at Harvard used an algorithm\cite{shakhnovich}
to design ten 48-mer proteins, and challenged a group at UC San Francisco
to fold these chains to their putative ground states. Because their
Monte Carlo simulations were unable to fold chains to energies as low as
the states they had designed, they made a wager with 
the UCSF group, that they would be unable to find conformations 
at the same energies
as their designed sequences. In fact, using their fast folding
algorithms, the UCSF group was able to find energies lower than Harvard's
putative ground states, and won a six pack of beer.

\begin{figure}[tbh]
\begin{center}
\
\psfig{file=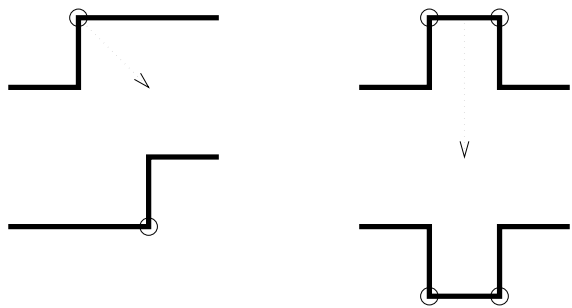,width=3in}
\end{center}
\caption{Illustration of local moves made in Monte Carlo simulations.}
\label{fig:local}
\end{figure}    

The usual lattice Monte Carlo move sets\cite{verdier,hilhorst} 
are illustrated in figure
\ref{fig:local}.  There are two basic moves, right angle moves, and top
hat moves.  However for moves to be successful, the chain has to move
into an unoccupied site. This is one reason why Monte Carlo is so slow
for proteins; they are typically very dense systems. If in a lattice
system all sites are occupied, it is impossible by these conventional
techniques to move a chain at all. Here we propose and implement a set
of long range moves that allow for fast motion of a chain {\em even when
all sites are occupied}. Other methods based on chain
growth algorithms\cite{rosenbluth} have been recently
introduced\cite{otoole,garel} and appear highly promising.
In particular, O'Toole and Panagiotopoulos tested their
method and showed that it could fold chains much longer
than is possible with short range Monte Carlo. They designed a
sequence to fold into an 80-mer four helix bundle and were
able to fold that sequence to a state close in energy and 
conformation, to that of their designed structure. 

We examine the same 48-mer sequences as above, testing
out local Monte-Carlo moves and these new long range moves.
We find the long range algorithm is an order of magnitude
faster for these sequences.

We finally consider the 80-mer four helix bundle mentioned
above and find that the our long range algorithm is able
to reach the putative ground state that it was designed 
to fold to. We also find that below the lowest energy state
found by O'Toole and Panagiotopoulos there are a variety
of other states with different structures often that are 
quite regular and sometimes rich in  beta sheets.

\section{Long Range Moves}

In this paper we concentrate on lattice
systems mainly because the problem of high density is most acute.
The idea of the long range moves proposed here is to allow chain
motion even at high densities. Conventional short range moves
require monomers to move from occupied to unoccupied positions.
At high density, such moves are mostly prohibited and the system
becomes log-jammed\cite{deutsch82}. 

Here we propose moves that allow the chain to change conformation
without any net change in the occupation of a site. We do so
by cutting the chain and re-splicing it in such a way  that
the final topology of the chain remains linear. 

\begin{figure}[tbh]
\begin{center}
\
\psfig{file=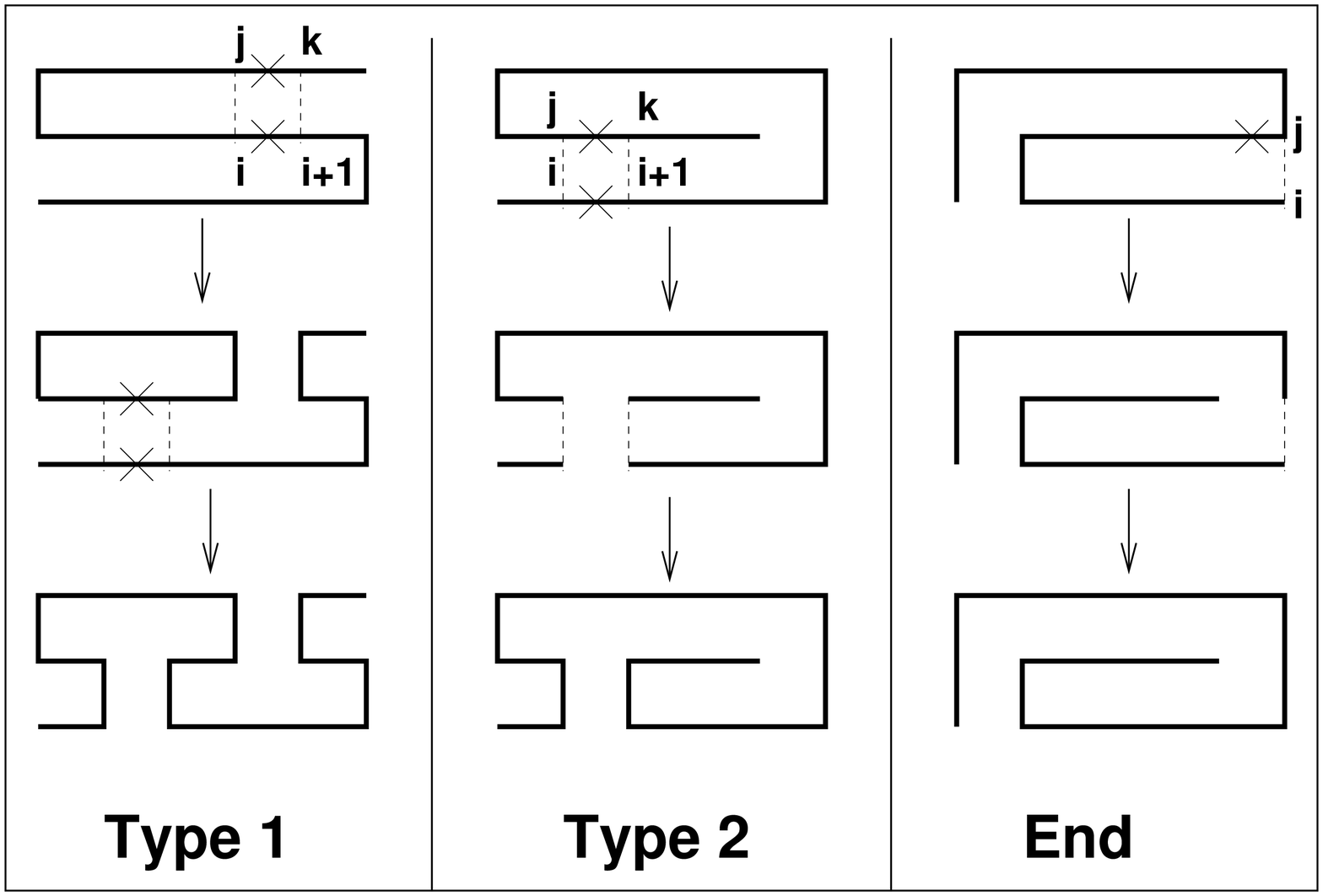,width=5in}
\end{center}
\caption{Illustration of long range moves.}
\label{fig:lrmoves}
\end{figure}

We will consider a model of a protein that lies on a cubic lattice,
with a lattice constant of unity. Double occupation of sites is
prohibited.
There are three basic types of move illustrated in figure \ref{fig:lrmoves}.
The first two, ``type 1" and ``type 2" have to do with motion of the
middle of the chain, the last type of move changes the position
of chain ends.

For type 1 and 2, the ith monomer is chosen at random. The displacement
vector between the monomer $i$ and  $i+1$ gives the local direction of the
chain. Next a random unit vector in a lattice direction is chosen and the neighbors
of monomers $i$ and $i+1$, in this direction, are determined. If these
two points are adjacent along the backbone then these
two strands both lie in the same direction, and so we continue, otherwise
the process starts again. Call the points opposite monomer $i$ and $i+1$
$j$ and $k$ respectively. If $k - j = +1$, the two strands are parallel,
otherwise they are anti-parallel. When they are parallel, we attempt
to  perform a type 1 move, if they are anti-parallel, we attempt a
type two move.

For a type 1 move, we then cut the links between $i$ and $i+1$, and between
$j$ and $k$. We then rejoin the chain, making links between $i$ and $j$ and between
$i+1$ and $k$. Now the chain is divided up into two separate pieces, a linear
portion and a circular piece. We then attempt to rejoin the circular
portion of chain to the linear one in a similar manner to what was described
above. In more detail, we try to find another position where strands on the circular
piece are parallel to those on the linear one. We do this by choosing another
strand at random and determining if along a randomly chosen direction, 
its neighboring strand is parallel or anti-parallel to it. If this is
the case, we reconnect similar to the manner stated above. Note the chain
has changed its configuration without needing to move into any unoccupied
sites.
 
Type 2 moves are less complex. When we rejoin the monomers
$i$, $i+1$, $j$ and $k$, as 
described in the last paragraph, the chain is in a linear topology,
so no more operations need be performed. 

For end moves, an end is chosen at random, and then a random unit vector
is chosen. The neighboring monomer in that direction, $j$, is found. 
Monomers $i$ and $j$ are joined together, and a link between $j$ and one
of its neighbors is cut as illustrated in figure \ref{fig:lrmoves}.

The time it takes to perform a move is linear in the length of the
chain, so this algorithm provides an efficient means to move a
chain even at high densities.

The drawback of long range moves is that one would expect a rather
low acceptance rate for them.  The difference between the initial and
final energy is typically quite large because in the regions that
have changed, neighboring monomers are completely different.
However if the changed regions involve only a few monomers, which can happen
for many configurations, the energy difference will also be small.
Furthermore, even if the acceptance rate is exponentially small in the
number of monomers that are altered, these
moves allow the molecule additional ways to move through phase space.
If the molecule is caught in a {\em cul de sac}, it can take a
time exponential in $\beta$ to tunnel out, and the introduction
of these moves allows it to jump to other branches that can
have lower energy.

\section{27-mer Cube}

To demonstrate the efficiency of this method, we show how it
can be used to find ground state energies of systems even
when all sites are occupied.  We consider the  27-mer cube
that has been extensively studied in previous work. We use
the long range moves describe above to determine the ground
states of different sequences. 

The HP model has two types of monomers, H (hydrophobic) and P (polar). 
If two monomers are nearest neighbors then they interact with an
energy of $-1$ if they are both H's, otherwise the interaction
energy is zero. The energy of a chain is therefore proportional
to the number of nearest neighbor contacts between H's. 

We generated 100 random sequences and attempted fold each one
using Monte Carlo to see if it reached the ground state. The
ground state was determined by using exact enumeration. 

The Monte Carlo method employed was the usual Metropolis algorithm.
Once a chain had changed to a new configuration, using the method
described in the last section, the difference between the new
and the old energies, $\Delta E$, was calculated and the move
was accepted with a probability $\exp (-\beta \Delta E)$ if the
$\Delta E > 0$, and was also accepted otherwise. In this
case $\beta = 2$. The chain
was always started in the same ``zig zag" conformations similar
to the initial shape of the type 1 move of figure \ref{fig:lrmoves}.

The method found the ground state energy for all 100 sequences.
The average number of trials it took was 4736.

The success of this method should be contrasted with the failure
of the usual move set employed in conventional Monte Carlo, see figure
\ref{fig:local}. In the latter case, the requirement
that the chain remain confined to a cube, prevents any moves from
being made. Therefore conventional Monte Carlo has difficulties
in this 27-mer example. These can be circumvented if we allow the
chain to take up configurations outside of the cubic boundary.
If we add a large enough attractive energy to all interactions,
then it guarantees that the ground state is a cube\cite{socci}. However the
addition of a sizeable additive term increases the relaxation time.

\section{HP 48-mer}

To test out the efficacy of the  long range moves for finding the 
ground states of longer chains, we used the published results of 
ref. \cite{whoops}, where the UCSF group had determined the exact 
ground states of 10 48-mer sequences. Unlike the previous
section, the chains are unconfined and can take on any conformation.
Therefore both short range and long range moves should work in principle,
for this problem. In fact the HP model allows for considerable
density fluctuations of the P monomers. This is because the 
P monomers are inert, as there is no interaction energy with their
nearest neighbors, of either monomer type. The ground state
conformations have a hydrophobic (H) core and a surface of polar
(P) monomers. The P monomers feel no attraction so that strings
of P monomers are often quite flexible. As a result we will see that
short range
moves can move P monomers around quite effectively. The hydrophobic
core of the molecule
is quite small, 24 monomers, and is irregularly shaped, so
that most of the H monomers also are quite mobile. Because
of this, it was quite surprising to learn that the Harvard group
had little success finding their real ground states with short
range moves.

We employed the technique of simulated annealing to increase the
efficiency of our search. The temperature of the system was 
decreased slowly. As the temperature was decreased, the 
time that the system equilibrated at one temperature was increased. 
During equilibration,
the lowest energy state found was recorded. When the temperature was
lowered, at the beginning of a new cycle,
the chain was started in this lowest energy state.

For purely short range moves, we were able to find the ground
state of all ten chains. The times found depended on the
random number seed, but the shortest time  observed was about a
minute on a 133Mhz pentium processor. The longest time was
about one day. The time, $t_i$, of the ith temperature cycle increased
exponentially as did $\beta$. 
It was chosen to be $t_{i+1} = 1.0004t_i$
and $\beta_{i+1} = 1.0002\beta_i$.

Long range moves were then added in. At the beginning of a temperature
cycle, short range moves were performed for
5 time steps\cite{note}. Then 240 long range moves were  attempted. Because a long
range move is less likely to be accepted than an individual 
short range move, $\beta$ was decreased by 75\% . After this
the molecule went through a cooling period of 2400 Monte Carlo steps.
The lowest energy was kept track of and was used at the start of
the next cycle. The cycle times for this case were chosen to be 
$t_{i+1} = 1.002t_i$ and $\beta_{i+1} = 1.001\beta_i$. 
The timings used were somewhat arbitrary but appeared to give
fast convergence. 

\begin{table}[tbh]
\begin{center}
\begin{tabular}{|c|c|c|c|}
\hline
Sequence & Local Moves & Long Range Moves & Ground State \\
\hline
1 & 5.38$\times 10^7$ & 2.47$\times 10^5$ & -32  \\
2 & 6.63$\times 10^7$ & 6.69$\times 10^6$ &  -34 \\
3 & 7.46$\times 10^6$ & 1.05$\times 10^6$ & -34 \\
4 & 6.43$\times 10^7$ & 1.18$\times 10^6$ &  -33  \\
5 & 6.04$\times 10^7$ &  5.70$\times 10^6$ & -32  \\
6 & 8.42$\times 10^7$ & 4.02$\times 10^6$ & -32  \\
7 & 6.91$\times 10^7$ & 2.95$\times 10^6$ & -32   \\
8 & 6.71$\times 10^7$ & 4.05$\times 10^6$ & -31 \\
9 & 1.51$\times 10^8$ & 1.74$\times 10^7$ & -34 \\
10 & 4.55$\times 10^6$ &  2.77$\times 10^6$ & -33 \\
\hline
\end{tabular}
\end{center}
\caption{Comparison of folding times for short and long range moves}
\label{tab:times}
\end{table}          

Table \ref{tab:times} tabulates the results. 
These were for individual runs, and many such
runs would be needed if one were interested in obtaining
the average folding time for each molecule.
In terms of operations,
one long range move is of order N (chain length)
short range moves, or one Monte Carlo step. Most
of the computer time is still spent making short range moves,
but the occasional long range move gives the molecule a large
jolt allowing it to efficiently explore more of its 
conformational space.

The long range moves improve the efficiency of the ground state search
by roughly an order of magnitude for these 48-mer sequences. As discussed
above, one would
expect the speed up to be more dramatic in cases where there was an attraction
between the P monomers. However only the HP model can be used as a
test of folding times for 48-mer chains, as it is only for this model that
ground state energies can rigorously be determined.

\section{80-mer Four-Helix Bundle}

\begin{figure}[tbh]
\begin{center}
\
\psfig{file=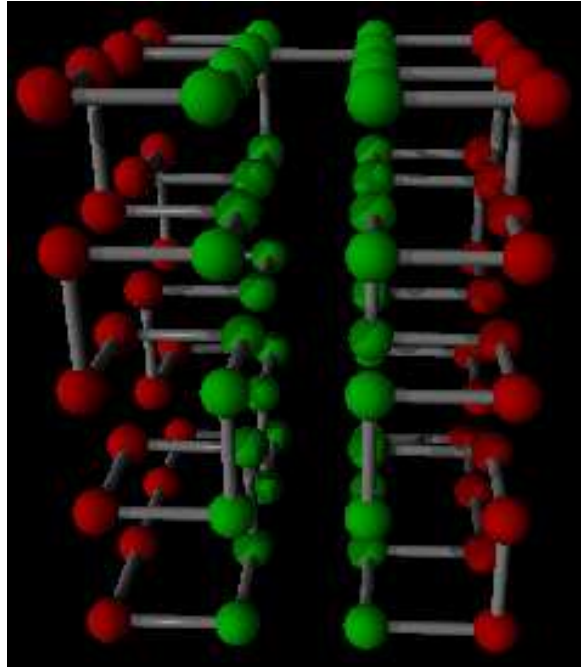,width=3in}
\end{center}
\caption{Four helix bundle designed 
by O'Toole and Panagiotopoulos to have an energy of -94.}
\label{fig:otoole}
\end{figure}

O'Toole and Panagiotopoulos\cite{otoole} tested the limits of their chain
growth algorithm by considering an 80-mer chain that
had two species of monomers. The interaction was
slightly different than the HP model, as both species
had an attraction of $-1$, but the there was no interaction
between two dislike monomers.  They  designed a
four-helix bundle that had an energy of $-94$ as pictured
in fig. \ref{fig:otoole}. They then used their fast growth
algorithm to attempt to fold the chain to this energy.
After growing 2.3 million chains, the lowest energy they
reached was -91. The conformations they found 
had a similar tertiary structure to the target conformation.

Using the same sequence\cite{sequence} the long range Monte
Carlo method was used to attempt to fold the chain. We used
a cycle similar to the 48-mer case described above. The
number of attempted long range moves was increased to 6400
and $\beta$ was decreased by 50\%. This was to ensure that
a reasonable number of long range moves were being made 
every cycle. It was found that it was most efficient to
give up after a certain amount of time and restart the
simulation with different random numbers. In this way
the entire simulation process was repeated 200 times. 
We were able to reach the putative ground state energy of -94.
This took about a week on a 133Mhz Pentium processor. The
number of operations appears to be roughly equivalent to
that used be O'Toole and Panagiotopoulos. However this
method was able to fold all the way to the putative
ground state. 

\begin{figure}[tbh]
\begin{center}
\
\psfig{file=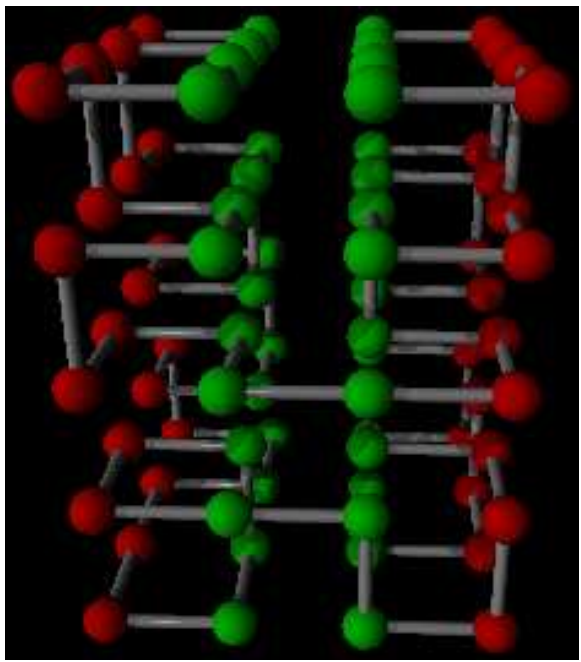,width=3in}
\end{center}
\caption{Alternative conformation found by the long range Monte
Carlo algorithm with an energy of -94 for the
80-mer considered by O'Toole and Panagiotopoulos.}
\label{fig:e94}
\end{figure}    

\begin{figure}[tbh]
\begin{center}
\
\psfig{file=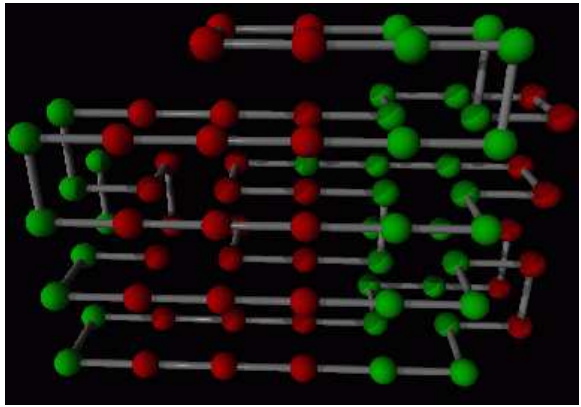,width=3in}
\end{center}
\caption{Conformation at an energy of -92 of the 80-mer considered
by O'Toole and Panagiotopoulos.}
\label{fig:e92}
\end{figure}

\begin{figure}[tbh]
\begin{center}
\
\psfig{file=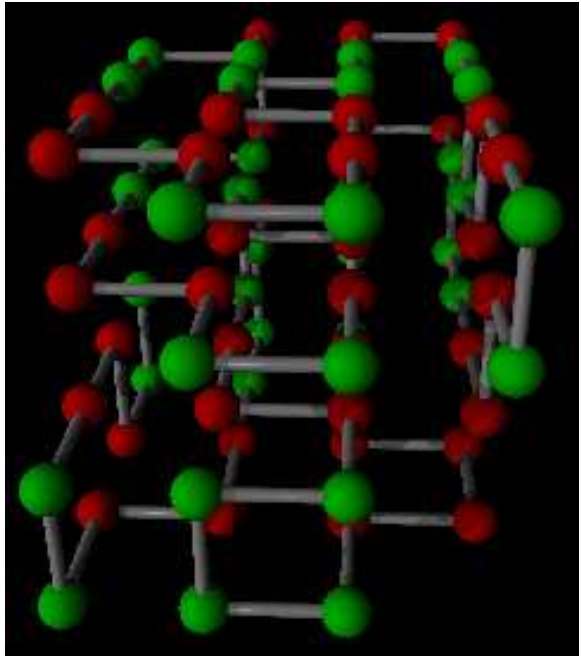,width=3in}
\end{center}
\caption{Conformation at an energy of -93 of the 80-mer considered
by O'Toole and Panagiotopoulos.}
\label{fig:e93}
\end{figure}

The conformations that were found are worthy of comment.
First the state found with energy of $-94$ was not
the same as the target structure. It is shown in figure \ref{fig:e94}.
It deviates substantially from the four-helix bundle.
There are two crossings between the left and the right sides
of the molecule instead of the one shown in figure \ref{fig:otoole}.
Second, the low lying excitations, with energies of $-92$
and $-93$ are often completely different than a four-helix
bundle. The structure shown in figure \ref{fig:e92} has an
energy $-92$ but is mostly made up of beta sheets. 
The conformation
in figure \ref{fig:e93} has an energy of $-93$ but is quite
different than any of the other figures.
This illustrates that the problem of design is quite
difficult. Negative design must be incorporated to
get rid of these competing structures and stabilize
the four-helix bundle. A framework\cite{KD} and method\cite{DK}
have recently been introduced for doing this.
It is also surprising how regular these alternate structures are
given the sequence was designed to fold to a completely different
conformation.

\section{Conclusion}
 
In this paper we have introduced and tested a set of long range
moves which allow for the equilibration of chains at high 
density. We have shown that this method works even for a lattice
model where all sites are occupied. 

For the 48-mer HP model
we find rather surprisingly, that even short range moves can
find the ground state for a set of sequences that had been
previously thought impossible to fold using only them.

Long range moves however considerably speed up folding times
and this algorithm was even able to find structures for 80-mer chains
that were previously unattainable even by very fast chain growth 
algorithms\cite{otoole}.

One has reason to hope that these long range moves might speed
up folding times even more for real proteins as there is far less
mobility of monomers on the surface, making short range moves
less effective than the 48-mer example. 
The same set of moves, with minor modifications can be used for more
realistic off-lattice simulations. In this case, it is hoped
that finding the tertiary structure will be sped up by the
introduction of this move set. 

\section{Acknowledgments}
The author thanks Tanya Kurosky, Eliot Dresselhaus, 
and Michele Vendruscolo for useful discussions.
This work is supported by NSF grant number DMR-9419362
and acknowledgment
is made to the Donors of the Petroleum Research Fund, administered
by the American Chemical Society for partial support of this research.


\begin{thebibliography}{99}
\bibitem{note} We use the usual convention that one Monte Carlo time step
is an attempted motion of every monomer of the chain.
\end{thebibliography}
\end{document}